%% file: ESUPPSustainability.tex
\documentclass{cernyrep}
\usepackage[utf8]{inputenc}
\usepackage[T1]{fontenc}
\usepackage{soul}
\usepackage{hyperref}
\usepackage{mmacells}
\usepackage{bbold}
\usepackage{slashed}
\input{ipython.tex}
\usepackage{listings}
\usepackage{fontawesome}
\usepackage[most]{tcolorbox}
\definecolor{light-gray}{gray}{0.9}
\usepackage{lineno}
\usepackage{mmacells}
\usepackage{listings}
\usepackage{comment}
\usepackage{hyperref}
\hypersetup{
    colorlinks,%
    citecolor=black,%
    filecolor=black,%
    linkcolor=black,%
    urlcolor=black,%
    pdftitle={Input to European Strategy Update for Particle Physics: Sustainability},%
}

\begin{document}
\setlength\parindent{10pt}
\title{Input to European Strategy Update for Particle Physics: Sustainability}

\author{
Veronique Boisvert\aff{1},
Daniel Britzger\aff{2},
Samuel Calvet\aff{3}, 
Yann Coadou\aff{4},
Caterina Doglioni\aff{9},
Julien Faivre\aff{5},
Patrick Koppenburg\aff{14},
Valerie S. Lang\aff{11},
Kristin Lohwasser\aff{6},
Zach Marshall\aff{7},
Rakhi Mahbubani\aff{8},
Peter Millington\aff{9},
Tomoko Muranaka\aff{16},
Karolos Potamianos\aff{15},
Ruth Pöttgen\aff{10},  
Hannah Wakeling\aff{12},
Efe Yazgan\aff{13}
}

\institute{
\naff{1}{Royal Holloway University of London, Egham, United Kingdom}
\naff{2}{Max Planck Institute for Physics, Garching, Germany}
\naff{3}{Université Clermont-Auvergne, CNRS, LPCA, Clermont-Ferrand, France}
\naff{4}{CPPM, CNRS/IN2P3, Aix Marseille Univ, Marseille, France}
\naff{9}{University of Manchester, Manchester, United Kingdom}
\naff{5}{Laboratoire de physique subatomique et de cosmologie, Grenoble, France}
\naff{14}{Nikhef, Amsterdam, Netherlands}
\naff{11}{Albert-Ludwigs-Universit\"at Freiburg, Freiburg, Germany}
\naff{6}{University of Sheffield, Sheffield, United Kingdom}
\naff{7}{Lawrence Berkeley National Laboratory, Berkeley, USA}
\naff{8}{Rudjer Boskovic Institute, Zagreb, Croatia}
\naff{16}{EPFL, Lausanne, Switzerland}
\naff{15}{University of Warwick, Coventry, United Kingdom}
\naff{10}{Lund University, Lund, Sweden}
\naff{12}{John Adams Institute, University of Oxford, United Kingdom}
\naff{13}{National Taiwan University, Taipei, Taiwan}
}
\vspace{-0.5em}
\begin{abstract}
Human activity continues to have an enormous negative impact on the ability of the planet to sustain human and other forms of life. Six out of the nine planetary boundaries have been crossed, a seventh is close to threshold. Prominent manifestations of this development are climate change caused by greenhouse gas emissions, as well as loss of biodiversity. In recognition of the urgency of these problems, several international agreements have been ratified to achieve net-zero emissions and to halt and reverse biodiversity loss. Significant reductions in emissions are required by 2030 to meet international climate targets. The field of particle physics has an obligation and an opportunity to contribute to such mitigation efforts and to avoid causing further harm. This document urges the European Strategy Update in Particle Physics to set a clear and bold mandate for embedding environmental sustainability throughout the future scientific programme, and advocates for a series of actions that will enable this.
\end{abstract}
\vspace{-0.5em}

\maketitle

\section{The case for sustainability in particle physics}

The scientific basis for climate change is well established~\cite{ManabeWetherald1967,Hasselmann1976,IPCC2021report}:\ The increase in CO$_2$ and other greenhouse gases (GHG) in the atmosphere interferes with Earth's natural carbon cycle and leads to changes in local and global temperatures~\cite{essd-16-2625-2024}. The year 2024 was the hottest, and the last ten years the warmest, on record~\cite{WMO2024,BerkeleyEarth2024,NASA2024,Copernicus2024}. Changes in ocean currents and melting of arctic ice are among the observations consistent with the predictions of climate science~\cite{Ripple2024}, and the framework of extreme event attribution~\cite{NAP21852} indicates that the current level of climate change can be fully attributed to human activity~\cite{IPCC2021reportSPM}. The resulting changes in weather patterns cause persistent draughts~\cite{draught1,draught2}, heavy storms~\cite{storm1,storm2}, and record levels of precipitation and floods~\cite{flood1,flood2}, leading to numerous deaths, severe disruption, and billions of euros in damage. Global warming of 2.7$^\circ$C or more above preindustrial levels will leave a third of the population outside the ``human climate niche'' for survival and activity~\cite{doi:10.1073/pnas.1910114117,lenton1}. 

The Paris Agreement~\cite{paris} aims to keep increases in global average temperatures well below 2$^\circ$C compared to preindustrial levels. Achieving its goals requires a reduction of net GHG emissions to zero by 2050. A significant amount of this reduction should be achieved by 2030. \emph{This is within the next five years, the same period targeted by the update to the European Strategy for Particle Physics.}

In addition to changes in global weather patterns, the variety and variability of life on Earth is threatened directly by human activities and indirectly through climate change.
This biodiversity is declining faster now than at any other time in human history, and current extinction rates are 100 to 1,000 times higher than the baseline and increasing~\cite{Ceballosbiodiversity,biodiv}. The degradation of Earth's ecosystems has a direct impact on humanity, and the cost of replacing their services with artificial solutions is high~\cite{ipbes_2019_6417333}.

It is clear that there are moral and practical imperatives for particle physics to act:
\begin{itemize}
    \item in climate mitigation:\ reducing energy and material resource consumption, GHG emissions, and other negative environmental impacts.
    \item in climate adaptation:\ readying our research infrastructure and practices for the manifold challenges posed by climate change.
\end{itemize}
Moreover, particle physics will be under increasing pressure to justify its impacts, and relying on grid decarbonization will not be enough to guarantee that large-scale  experiments will be deemed viable.

The emissions from particle physics are sizeable. The 2022 CERN emissions were 361,000~tCO$_2$e (tonnes of CO$_2$ equivalent)~\cite{CERNEnvironment2023}. These emissions can be compared with those of a developed country like Liechtenstein (with a population of about 40,000), which emitted 169,000 tCO$_2$e~\cite{WIDannualCo2e} in 2022\footnote{CERN's emissions include all scope 1, scope 2 and scope 3 emissions with the scope 2 emissions calculated over a reference timescale of 100 years. Country GHG emissions include carbon dioxide, methane and nitrous oxide from all sources, including land-use change. They are measured in tonnes of carbon dioxide-equivalents over a 100-year timescale.}~\cite{wikiLiechtenstein}. Not counting the embodied carbon in the civil engineering and machines themselves, the GHG emissions due to 2.5 minutes of CERN operations (1 tCO$_2$e) equates to just over 1 year of per capita beef consumption or 1.5 years of per capita dairy consumption~\cite{OWIDfoodaverage, OWIDmilkaverage, OWIDfoodemissions}, 3.5 years of per capita clothes shopping~\cite{TextileEmissions}\footnote{This is based on average consumption in the EU.}, 6,000~km of car travel~\cite{Ritchie_2023}, or 10 years of per capita emissions in Rwanda~\cite{RwandaEmissions}, making clear the insidiousness of climate injustice. Particle physics has made substantial progress in acknowledging the importance of sustainability and estimating GHG emissions and other environmental impacts (see, e.g.,~\cite{Bloom:2022gux, Banerjee:2023avd, yHEP2023a, yHEP2023b, Bloom:2024ixe, Lang:2024wpo, Saha:2024prx}, the CERN Environment reports~\cite{CERNEnvironment2020, CERNEnvironment2021, CERNEnvironment2023}, and the forthcoming Lab Directors Group guidelines on the Sustainable Assessment of Future Colliders~\cite{LDGWork}).
 
This document aims to complement the emphasis given to sustainability in other inputs to the European Strategy in Particle Physics and to urge the Update to give the strongest mandate to embedding environmental sustainability throughout European particle physics:\ from decision-making through to the design, construction, operation and decommissioning of future experimental infrastructure. Doing so presents both challenges and opportunities, and these are identified below. A list of key measures was identified by the community, and these are provided to inform a significant expansion of the European Strategy Update's emphasis on sustainability.

\section{Challenges and opportunities}

Tackling the climate emergency and the degradation of the world’s ecosystems is a systems problem. Moreover, it is a ``wicked’’ problem~\cite{Rittel1973}, that is not immune to economics, politics, geopolitics, sociology, or psychology, and progress requires systemic organizational transformation, alongside behaviour and culture change. However, there are also opportunities that arise from the need to transition the system to a more sustainable mode of operation. 

For particle physics, the transition is a problem that relies heavily on the provision of resources and funding, and the actions of many external agents: funders, governments, industry partners, and the up- and downstream supply chains. Moreover, the key drivers for particle physics {---} scientific questions, but also political and sociological aspects as well as available funding {---} are very different to other sectors, where consumer pressure can be a powerful force. Even so, the social licence to operate could be as easily lost for selected areas of science. Notwithstanding increasing pressure from governments and funders, much of the impetus to improve the environmental sustainability of particle physics has and will continue to come from within the scientific community. This can place environmental sustainability in direct competition with other priorities and sometimes perverse incentives, primarily driven by external factors (e.g., performance metrics that reward frequent travel). Changes in these incentives can help to avoid rebound effects, where large demonstrative achievements in sustainability trigger, and are offset by, increased consumption of resources. Nevertheless, environmental sustainability can be in direct tension with other progressive community-driven actions, e.g., to address barriers to equity, diversity, inclusivity, and accessibility. 

There are manifold real and perceived implications of changes made to improve environmental sustainability. These legitimate concerns must be mitigated against and they commonly relate to:
\begin{itemize}
\item reduced scientific output, and subsequent loss of competitive edge on the level of individuals, collaborations, institutions, nations, and regions;
\item less effective dissemination of scientific results;
\item decreased international cohesion of scientific effort;
\item changes in career and professional development, and the need for upskilling;
\item disproportionate impacts on particular groups, e.g., those who are geographically (or otherwise) isolated, those with caring responsibilities, or those with health conditions or impairments.
\end{itemize}

Taking a systems approach~\cite{systemsapproach} to this challenge can, however, allow capitalization on co-benefits: for equity, inclusivity, diversity and accessibility, for helping to drive the Green industrial revolution, and for helping to strengthen international relationships.  This includes consideration of the optimal siting of research infrastructure, which can have additional benefits for reducing geographic disparities. It offers opportunities to proactively re-engineer the way that we do our science, to collaborate far outside the usual boundaries of our disciplines, and to engage directly with economists, political scientists, and social scientists~\cite{creativedestruction}.

\section{Sustainability actions for the European Strategy for Particle Physics}

\textbf{Long-term environmental sustainability must be one of the guiding principles of scientific strategy}, and we advocate that: 

\begin{itemize}\addtolength{\itemsep}{0.4\baselineskip}

\item[$\bullet$] \textbf{Future particle physics projects, including next energy-frontier machines, be evaluated on adherence to principles of environmental sustainability from design to disposal, based on comprehensive Life Cycle Assessments (LCAs).}

\vspace{.2cm}

Particle physics projects often require a significant amount of resources in terms of materials, funding, and person power. Proposals for future experiments must include rigorous environmental impact studies following standard LCA practices, from cradle to grave, including construction, embedded impact and operations, and decommissioning, not limiting these impacts to GHG emissions.  With an envisaged start of operations in 2045--2050, the timeline of the future flagship project overlaps with that of the net-zero pledges by EU countries.  LCAs are underway to assess the environmental impact of proposed projects like the FCC~\cite{Benedikt:2928194}, and ILC and CLIC~\cite{ARUPLCA}.  The results must be used to refine the project design for mitigation of assessed impacts, and in order to achieve net-zero emissions using existing and emerging technologies, such as high-temperature superconductors, energy recovery linacs, and green cement. 
Implementing a modest delay in the schedule of the next generation of colliders would allow for the implementation of mitigation measures, while simultaneously providing time to fully exploit the physics potential of the HL-LHC. Aligning cutting-edge research with the climate concerns emphasized by Early Career Researchers~\cite{ECR,arling2025earlycareerresearcherinput} will help secure future talent.

\vspace{.2cm}

Vital to this end, given the bespoke nature of the materials and equipment used in particle physics experiments, is the allocation of specific human and financial resources to compiling relevant LCA libraries. Engaging with other research fields is crucial, to encourage R\&D programmes to effectively minimize emissions at every stage of the project using LCAs, e.g., in relation to green cement for construction, flexibility in electricity consumption optimized for renewable energy sources for operation, and recycling capabilities and material usage for decommissioning.
Training in LCA techniques and sustainability issues should be encouraged for the whole particle physics community, and career paths should be developed for particle physicists who wish to become experts in the cross-field issue of the sustainability of large-scale scientific experiments.

\item[$\bullet$] \textbf{All institutions involved in particle physics, including  laboratories and universities, publish environmental reports at least every two years and use these to define ambitious emissions targets that they hold themselves accountable to.
}

\vspace{.2cm}
Reports should contain all relevant quantitative and qualitative information, such as GHG emissions, total energy use, water use, waste output, and biodiversity initiatives.  The Laboratory Directors Group (LDG) recommendations for environmental reporting should be followed when they become available, and meanwhile the CERN environmental reports~\cite{CERNEnvironment2020,CERNEnvironment2021,CERNEnvironment2023} could be used as a template. Environmental targets, such as emissions and energy reductions, must be developed based on the institute’s emissions profile, and with the aim to reach net zero in compliance with national and European targets.  These should be made publicly available.

\item[$\bullet$] \textbf{CERN, with its leading role in particle physics, hold itself to emissions reduction beyond its current targets, demonstrating the field’s commitment to realizing net zero on a timescale compatible with the goals of the Paris Agreement, and setting the standard for basic scientific research worldwide.}
 
 \vspace{.2cm}
 Urgent carbon emission reductions are essential to mitigating climate change. While CERN's emissions are not included in national UNFCCC inventories, its environmental reports align with reporting requirements but lack a clear justification for reduction targets. CERN currently aims for a 50\% emissions reduction by 2030, with respect to 2018 levels, commensurate with Swiss and French targets but 
 falling significantly short of the EU target recommended by Climate Action Tracker~\cite{CAT} for compatibility with the Paris Agreement's 1.5$^\degree$C limit. To signal a strong commitment to mitigating climate change and to demonstrate leadership in sustainability, CERN should surpass national targets and strive for at least a 60\% reduction by 2030 with respect to 2018 emissions.

\item[$\bullet$] \textbf{Laboratories that host particle physics experiments hold themselves accountable to and go beyond the environmental pledges of their hosting countries, including those from which they may be exempt, e.g., the upcoming ban on the emission of fluorinated gases (F-gases).}

\vspace{.2cm}
A significant part of CERN's direct GHG emissions is due to the leakage of F-gases with high global warming potential from particle detectors, where they are used for cooling and particle identification~\cite{CERNEnvironment2023}.  There are already significant and crucial R\&D activities aiming at the complete replacement of F-gases for new equipment, and the recuperation and recirculation of gases, and prevention of detector leaks for current detectors. Investment into these R\&D activities should increase. There should also be continued R\&D aimed at replacing gas detectors with alternative detectors for future experiments, in case no suitable gases are found that satisfy all requirements. Building on CERN’s first F-gas policy~\cite{CernFgas}, a similar phase-out and veto on future usage must be developed for other environmentally harmful materials such as forever chemicals (PFAS). Clear policies regarding the usage and phase-out must be developed by every major particle physics laboratory in Europe --- and worldwide --- as well as at universities and research institutions participating in experiments at the laboratories.

\item[$\bullet$] \textbf{CERN, national laboratories, funding agencies, and institutions encourage, reward, and provide opportunities for their scientists and engineers to contribute to addressing and communicating the environmental crisis.}

\vspace{.2cm}
Mitigating climate change is a global challenge that will require expertise from across diverse research fields, and particle physics has a responsibility to contribute. This responsibility extends beyond R\&D directly linked to the core physics programme, to education and public engagement on climate change, as well as climate technology and economy.  Knowledge exchange initiatives like the CERN-led CIPEA~\cite{CIPEA} are an effective way to harness field-specific skills for the net-zero agenda, and should be fostered and expanded.  To 
ensure strong engagement, it is essential that institutions provide specific training, funding opportunities, and recognition of such efforts in the context of career progression.

\item[$\bullet$] \textbf{{Negative environmental impacts of particle physics computing be quantified and reduced by integrating sustainability into hardware and software practices.}}

\vspace{.2cm}
The computing demands of particle physics continue to increase, resulting in significant and growing GHG emissions, not just due to energy consumption of computing infrastructure, but also hardware production.  Integrating sustainability into computing practices will allow for a significant reduction of emissions without impacting scientific progress. These considerations are not unique to particle physics, and close collaboration with groups already confronting the issue of sustainability in computing should be pursued. Computing policies for institutions and data centres should be carefully examined for opportunities to reduce their environmental impacts, and regularly updated.  
Users and software developers should receive training in mitigation of emissions through good software development practices and benchmarking, and other environmental impacts of computing. The WLCG, based at CERN, is well placed to coordinate community efforts around sustainability in computing, and could provide a forum in which sites can share best practices.

\item[$\bullet$] \textbf{CERN, national laboratories, funding agencies, and institutions develop flexible policies to enable and encourage environmentally friendly business travel, commuting, and methods of collaboration.}

\vspace{.2cm}
Mobility emissions are dominated by air travel --- one intercontinental return flight can sum to over half annual per capita emissions in Germany, already significantly higher than is compatible with net-zero emissions by 2050.  The emerging transition away from a ``flyout culture''~\cite{Poggioli2022} should be encouraged and incentivized in institutional travel policy, and facilitated by collaborations and in conferences by judicious use of digital tools and other initiatives~\cite{noflytoolkit}.   The choice of location of meetings, workshops and conferences should take into account the travel emissions of the participants, and multiple hubs should be considered.  Care must be taken to design policies and make these changes so as also to promote equity, diversity, inclusivity, and accessibility in the field.

\item[$\bullet$] \textbf{CERN, national laboratories, funding agencies, and institutions enable and encourage practices that minimize food-related GHG emissions and other associated negative environmental impacts.} 

\vspace{.2cm}
While food emissions are smaller than other emissions associated directly with particle physics, the food sector accounts for 25--30\% of global GHG emissions~\cite{Ritchie_2021}, and a quarter of all food produced is thrown away~\cite{Ritchie_2020}.  Simplifying the provision of and incentivizing environmentally friendly, largely plant-based, catering are crucial steps for research institutions to reduce GHG emissions in this area.  Some institutions and research communities have already made this shift, known to have reduced emissions and impact on the environment, and additional health benefits~\cite{WILLETT2019447}.  As societies move ever closer to net zero, all sources of emissions will need to be examined.

\item[$\bullet$] \textbf{Existing research sites be managed and future sites designed to enhance habitat diversity, halt local biodiversity loss, minimize water use, and encourage natural water cycles.}

\vspace{.2cm}
Current and future sites of particle physics research should exemplify biodiversity development~\cite{COP15}. The introduction of invasive alien species should be avoided, and experts in biodiversity should be consulted when designing and updating buildings, streets, or underground networks, and to reduce sources of disturbance, such as physical obstacles, traps, light, and noise. Soil and vegetation should be used to manage pluvial water to reduce ground movements, flood and drought hazards, and to minimize the rainwater efflux from the site. Water storage in soil allows plants to provide efficient vegetal air-conditioning through evapotranspiration, and the use of impermeable materials should be avoided, e.g., for roads and car parks.
\end{itemize}
\vspace{-1.3em}

\section{Conclusion}
The climate emergency, biodiversity loss, and chemical pollution of the environment present very real challenges and very real opportunities for particle physics. To address these challenges and to capitalize on these opportunities requires a large-scale, systematic and coordinated response that can be set firmly in motion by a clear and bold mandate from the European Strategy Update for Particle Physics.  This mandate needs to be far reaching, and it needs to filter through to every aspect of the way that particle physics is done:\ from the infrastructure that enables it, to the very core of how the international particle physics community operates.

\vspace{-0.6em}

\bibliographystyle{JHEP}
\bibliography{ESUPPSustainability}

\end{document}

%% file: ipython.tex
\definecolor{maroon}{cmyk}{0, 0.87, 0.68, 0.32}
\definecolor{halfgray}{gray}{0.55}
\definecolor{ipython_frame}{RGB}{207, 207, 207}
\definecolor{ipython_bg}{RGB}{247, 247, 247}
\definecolor{ipython_red}{RGB}{186, 33, 33}
\definecolor{ipython_green}{RGB}{0, 128, 0}
\definecolor{ipython_cyan}{RGB}{64, 128, 128}
\definecolor{ipython_purple}{RGB}{170, 34, 255}

\usepackage{listings}
\lstdefinelanguage{iPython}{
    morekeywords={access,and,break,class,continue,def,del,elif,else,except,exec,finally,for,from,global,if,import,in,is,lambda,not,or,pass,print,raise,return,try,while},%
    morekeywords=[2]{abs,all,any,basestring,bin,bool,bytearray,callable,chr,classmethod,cmp,compile,complex,delattr,dict,dir,divmod,enumerate,eval,execfile,file,filter,float,format,frozenset,getattr,globals,hasattr,hash,help,hex,id,input,int,isinstance,issubclass,iter,len,list,locals,long,map,max,memoryview,min,next,object,oct,open,ord,pow,property,range,raw_input,reduce,reload,repr,reversed,round,set,setattr,slice,sorted,staticmethod,str,sum,super,tuple,type,unichr,unicode,vars,xrange,zip,apply,buffer,coerce,intern},%
    sensitive=true,%
    morecomment=[l]\#,%
    morestring=[b]',%
    morestring=[b]",%
    morestring=[s]{'''}{'''},
    morestring=[s]{"""}{"""},
    morestring=[s]{r'}{'},
    morestring=[s]{r"}{"},%
    morestring=[s]{r'''}{'''},%
    morestring=[s]{r"""}{"""},%
    morestring=[s]{u'}{'},
    morestring=[s]{u"}{"},%
    morestring=[s]{u'''}{'''},%
    morestring=[s]{u"""}{"""},%
    literate=
    {á}{{\'a}}1 {é}{{\'e}}1 {í}{{\'i}}1 {ó}{{\'o}}1 {ú}{{\'u}}1
    {Á}{{\'A}}1 {É}{{\'E}}1 {Í}{{\'I}}1 {Ó}{{\'O}}1 {Ú}{{\'U}}1
    {à}{{\`a}}1 {è}{{\`e}}1 {ì}{{\`i}}1 {ò}{{\`o}}1 {ù}{{\`u}}1
    {À}{{\`A}}1 {È}{{\'E}}1 {Ì}{{\`I}}1 {Ò}{{\`O}}1 {Ù}{{\`U}}1
    {ä}{{\"a}}1 {ë}{{\"e}}1 {ï}{{\"i}}1 {ö}{{\"o}}1 {ü}{{\"u}}1
    {Ä}{{\"A}}1 {Ë}{{\"E}}1 {Ï}{{\"I}}1 {Ö}{{\"O}}1 {Ü}{{\"U}}1
    {â}{{\^a}}1 {ê}{{\^e}}1 {î}{{\^i}}1 {ô}{{\^o}}1 {û}{{\^u}}1
    {Â}{{\^A}}1 {Ê}{{\^E}}1 {Î}{{\^I}}1 {Ô}{{\^O}}1 {Û}{{\^U}}1
    {œ}{{\oe}}1 {Œ}{{\OE}}1 {æ}{{\ae}}1 {Æ}{{\AE}}1 {ß}{{\ss}}1
    {ç}{{\c c}}1 {Ç}{{\c C}}1 {ø}{{\o}}1 {å}{{\r a}}1 {Å}{{\r A}}1
    {€}{{\EUR}}1 {£}{{\pounds}}1
    {^}{{{\color{ipython_purple}\^{}}}}1
    {=}{{{\color{ipython_purple}=}}}1
    {+}{{{\color{ipython_purple}+}}}1
    {*}{{{\color{ipython_purple}$^\ast$}}}1
    {/}{{{\color{ipython_purple}/}}}1
    {+=}{{{+=}}}1
    {-=}{{{-=}}}1
    {*=}{{{$^\ast$=}}}1
    {/=}{{{/=}}}1,
    literate=
    *{-}{{{\color{ipython_purple}-}}}1
     {?}{{{\color{ipython_purple}?}}}1,
    identifierstyle=\color{black}\ttfamily,
    commentstyle=\color{ipython_cyan}\ttfamily,
    stringstyle=\color{ipython_red}\ttfamily,
    keepspaces=true,
    showspaces=false,
    showstringspaces=false,
    rulecolor=\color{ipython_frame},
    frame=single,
    frameround={t}{t}{t}{t},
    framexleftmargin=6mm,
    numbers=left,
    numberstyle=\tiny\color{halfgray},
    backgroundcolor=\color{ipython_bg},
    basicstyle=\footnotesize\ttfamily,
    keywordstyle=\color{ipython_green}\ttfamily,
    aboveskip=1.2em,
    belowskip=1.2em,
}